\documentclass[numberedappendix,apj,twocolumn,letter]{emulateapj}

\usepackage{amssymb}
\usepackage{amsmath}
\newcommand{\be}{\begin{equation}}
\newcommand{\ee}{\end{equation}}

\shorttitle{The GRB-LBG connection} 
\shortauthors{Trenti et al.}

\begin{document}


\title{Gamma-Ray-Burst Host Galaxy Surveys at Redshift $z\gtrsim4$:
  Probes of Star Formation Rate and Cosmic Reionization}

\author{Michele Trenti\altaffilmark{1,2,\dag},
Rosalba Perna\altaffilmark{3},
Emily M. Levesque\altaffilmark{2,\ddag},
J. Michael Shull\altaffilmark{2},
John T. Stocke\altaffilmark{2}
}
\altaffiltext{1}{Institute of Astronomy, University of  Cambridge, Madingley Road, Cambridge,  CB3 0HA, United Kingdom}
\altaffiltext{2}{CASA, Department of Astrophysical and Planetary Sciences, University of Colorado, 389-UCB, Boulder, CO 80309 USA}
\altaffiltext{3}{JILA, Department of Astrophysical and Planetary Sciences, University of Colorado, 389-UCB, Boulder, CO 80309 USA}
\altaffiltext{\dag}{Kavli Fellow}
\altaffiltext{\ddag}{Einstein Fellow}
\email{trenti@ast.cam.ac.uk} 

%

\begin{abstract}
  
  Measuring the star formation rate (SFR) at high redshift is crucial
  for understanding cosmic reionization and galaxy formation. Two
  common complementary approaches are Lyman-Break-Galaxy (LBG) surveys
  for large samples and Gamma-Ray-Burst (GRB) observations for
  sensitivity to SFR in small galaxies. The $z\gtrsim4$ GRB-inferred
  SFR is higher than the LBG rate, but this difference is difficult to
  understand, as both methods rely on several modeling assumptions.
  Using a physically motivated galaxy luminosity function model, with
  star formation in dark-matter halos with virial temperature
  $T_{vir}\gtrsim2\times10^{4}~\mathrm{K}$
  ($M_{DM}\gtrsim2\times10^{8}~\mathrm{M_{\sun}}$), we show that GRB
  and LBG-derived SFRs are consistent if GRBs extend to faint galaxies
  ($M_{AB}\lesssim-11$). To test star formation below the detection
  limit $L_{lim}\sim0.05L^*_{z=3}$ of LBG surveys, we propose to
  measure the fraction $f_{det}(L>L_{lim},z)$ of GRB hosts with
  $L>L_{lim}$. This fraction quantifies the missing star formation
  fraction in LBG surveys, constraining the mass-suppression scale for
  galaxy formation, with weak dependence on modeling
  assumptions. Because $f_{det}(L>L_{lim},z)$ corresponds to the ratio
  of star formation rates derived from LBG and GRB surveys, if these
  estimators are unbiased, measuring $f_{det}(L>L_{lim},z)$ also
  constrains the redshift evolution of the GRB production rate per
  unit mass of star formation. Our analysis predicts significant
  success for GRB host detections at $z\sim5$ with
  $f_{det}(L>L_{lim},z)\sim0.4$, but rarer detections at $z>6$. By
  analyzing the upper limits on host-galaxy
  luminosities of six $z>5$ GRBs from literature data, we infer
  that galaxies with $M_{AB}>-15$ were present at $z>5$ at $95\%$
  confidence, demonstrating the key role played by very faint galaxies
  during reionization.

\end{abstract}


\section{Introduction}\label{sec:intro}

The early stages of star formation at redshift $z\gtrsim4$ are being
probed by a growing number of observations. Lyman Break Galaxy (LBG)
surveys are taking advantage of the Wide Field Camera 3 (WFC3) on the
{\it{Hubble Space Telescope (HST)}}, pushing the frontier of galaxy detection to
$z\sim10$ \citep{bouwens11_nat,oesch11}. There are now samples of
several thousand galaxies at $z\lesssim 6$ \citep{bouwens07}, and more
than 100 galaxies at $z\sim7-8$
\citep{bouwens11,bunker10,finkelstein10,trenti11a,trenti11b}. Narrow-band
observations discovered large samples of Ly$\alpha$ emitter galaxies
at $z\sim5-7$ \citep{shimasaku06,ouchi10,ota10}. Large-area,
ground-based surveys are detecting QSOs at $z\sim6-7$
\citep{fan06,mortlock11}, shedding light on the formation of the first
super-massive black holes. Gamma-Ray-Burst (GRB) observations with
{\it{SWIFT}} have detected the object with the highest spectroscopically
robust redshift ($z=8.2$; \citealt{tanvir09,salvaterra09}) and a
photometric candidate at $z\sim9.4$ \citep{cucchiara11}, providing
an independent probe of the star formation rate (SFR) during the epoch
of hydrogen reionization \citep{kistler09,virgili11,robertson12}. At
the same time, numerical simulations and theoretical modeling are
addressing the formation of stars and galaxies within the first
billion years with increasingly successful predictions
\citep{trenti09b,trenti10a,lacey11,dijkstra11,finlator11,jaacks11}.

Despite this progress, many open questions on early star formation
remain. In fact, we do not know with confidence when ionization was
completed or sources responsible for it. The
WMAP-inferred epoch of cosmic reionization ($z\sim10.6\pm1.2$;
\citealt{komatsu11}) is somewhat in tension with galaxy observations
that suggest that reionization extended to lower redshift
($z\sim6-7$), based on evolution of the LAE luminosity function
\citep{ouchi10,ota10}. In addition, the photon budget from observed
galaxies at $z\sim6-10$ falls short of that required to produce the
optical depth to reionization measured by WMAP
\citep{bolton07,shull08,trenti10a,shull11}. A possible solution is
that there is significant star formation in small galaxies with
luminosity below $M_{AB}\sim-18$, the current limit for deep LBG
surveys at $z\gtrsim6$. The faint-end slope of the galaxy
luminosity function (LF) is very steep, $\phi(L)\propto
(L/L^*)^{\alpha}\exp(-L/L_*)$ in the \citet{schechter76} form, with
$\alpha\sim-2$. Because observed LBGs live in massive dark-matter halos
($M_{DM}\gtrsim10^{10}~\mathrm{M_{\sun}}$), most of the $z\gtrsim6$
star formation could take place in smaller halos that host dwarf-like
galaxies, with luminosity in the range $-18\lesssim M_{AB}\lesssim-10$.
This corresponds to the limit of Ly$\alpha$-cooling halos (virial
temperature $T_{vir}\gtrsim2\times10^{4}~\mathrm{K}$ and
$M_{DM}\gtrsim2\times10^8~\mathrm{M_{\sun}}$), in broad agreement with
theoretical modeling \citep{trenti10a}.

The prediction that faint dwarfs are the main agents of reionization
is difficult to test observationally. Even the {\it{James Webb Space
    Telescope (JWST)}} will be unable to reach the required sensitivity, as
it will only improve the sensitivity by $\Delta M_{AB}\sim2$ compared
to HST/WFC3 \citep{gardner06}.  In this respect, GRB observations
offer an independent probe of the total SFR, unlimited by the
faintness of the host galaxy and well suited to investigate star
formation during the epoch of reionization. Indeed, GRB rates suggest
that the total SFR at $z\gtrsim5$ is larger than that inferred from
LBG observations and the difference arises
because most of the star formation happens in small galaxies
\citep{kistler09,robertson12}. Still, deriving the SFR from GRB
observations is challenging, both because of the small number ($19$)
of $z>4$ GRB events with spectroscopically confirmed redshifts and
because detailed modeling is required to translate the GRB rates into
a SFR. In this respect, both selection and follow-up biases might be
important. For example, GRBs with dusty sight lines will be
under-represented \citep{robertson12}, and possibly GRBs with faint
afterglows as well, depending on the inhomogeneous follow-up data available.

Following \citet{kistler09}, the typical modeling assumes that
the comoving GRB rate is related to the SFR by:
\begin{equation}\label{eq:GRB_rate}
\dot{n}_{GRB}(z)=\varepsilon(z)\times\dot{\rho}_*(z),
\end{equation}
where $\varepsilon(z)$ is the efficiency of GRB production per unit
stellar mass, with a redshift dependence that can be used to model
biases in the relation between $\dot{\rho}_*$ and $\dot{n}_{GRB}$
(e.g., \citealt{robertson12}). This quantity is often modeled as:
\begin{equation}\label{eq:varepsilon}
\varepsilon(z)=\varepsilon_0 (1+z)^{\beta},
\end{equation}
with $\beta \approx -1.2$ derived empirically at $z<4$ and
extrapolated to higher redshift to infer $\dot{\rho}_*(z)$
(\citealt{virgili11,robertson12}).

To complement existing studies, we propose a novel idea for testing
the relation between SFRs from GRB and LBG surveys and for
investigating whether $\beta$ remains constant at $z>4$. We suggest
measuring the fraction of GRB hosts detected from observations
reaching the same magnitude limit as the typical LBG galaxy search. In
Section~\ref{sec:model}, we show that this fraction quantifies the
amount of star formation missed in LBG surveys, elucidating the
minimum luminosity (and halo masses) of galaxies in the epoch of
reionization. Measuring GRB-host detection efficiency at different
redshifts also constrains $\beta$. In Section~\ref{sec:predictions} we
consider the prospects for carrying out our proposed measure and show
a preliminary application of our method to recent {\emph{HST}} observations by
\citet{tanvir12}. Section~\ref{sec:con} summarizes our conclusions.

\section{The GRB-host detection fraction as probe of star formation
}\label{sec:model}

Both LBG and GRB surveys provide estimates of the SFR: from the
observed galaxy light in the first case, and from the observed GRB
rate in the second. Both approaches rely on modeling assumptions, such
as completeness estimates, dust extinction/obscuration, initial mass
function (IMF), age and metallicity of the stellar
populations. Therefore, a discrepancy between the two estimates can
either have a physical or a systematic origin. In
Figure~\ref{fig:sfr}, we report the latest determination of the
LBG-SFR from \citet{bouwens11} and the GRB-inferred SFR from
\citet{kistler09} and \citet{robertson12}. The dust-corrected LBG SFR
has been derived for galaxies with $L>0.05L_{z=3}^{*}$
($M_{AB}<-17.7$), the approximate limit of the \citet{bouwens11}
observations. It is immediately clear that the GRB SFR is
systematically above the LBG SFR. A possible explanation for the
difference is that there is significant star formation in galaxies
with $M_{AB}>-17.7$ \citep{kistler09}. To illustrate this, we plot in
Figure~\ref{fig:sfr} the SFR inferred from the luminosity function
model of \citet{trenti10a}, calibrated on the $z=6$ luminosity
function (LF) with the latest dust correction applied by
\citet{bouwens11}. This model is based on relating the evolution of
the galaxy luminosity function to that of the dark-matter halo mass
function via a modified abundance matching and results in LBG LFs that
are close to a Schechter form and match the observation well (see
\citealt{trenti10a}). We show two model predictions with different
assumptions on the luminosity below which galaxy formation is
suppressed: $M_{suppr}=-17.7$ (solid-black line) and $M_{suppr}=-11$
(dashed-green line). The higher limit corresponds to the limit of the
\citet{bouwens11} observations, demonstrating that the model
successfully reproduces the evolution of the LBG LF from $z=5$ to
$z=10$ (the model assumptions are not appropriate for
$z\lesssim5$). The lower limit assumes that star formation proceeds in
DM halos with smaller mass compared to that of the {\it{Hubble}}
ultradeep field galaxies, down to the limit of Ly$\alpha$ cooling
($T_{vir}\sim2\times10^4~\mathrm{K}$). In most models of galaxy
formation, these small halos are capable of cooling and forming stars,
which are included in the GRB-derived SFR. The model prediction
(Figure~\ref{fig:sfr}) is in agreement with the data within their
uncertainty.

Another explanation for the difference in the observed SFRs is the
possibility of systematic errors. For example, the redshift
evolution of the GRB production efficiency $\varepsilon(z)$ may
differ from the $(1+z)^{\beta}$ derived from $z<4$ data. To
investigate which of the two hypotheses is correct, we introduce the
idea of using the information contained in the fraction of GRB hosts
detected at a given redshift.

To present our framework, we assume for simplicity that the stellar
mass-to-light ratio does not depend on galaxy luminosity, and that the
GRB rate is proportional to the SFR, with no bias depending on
host-galaxy luminosity. We assume that galaxy properties such as
metallicity, dust content and IMF do not depend on $L$ at a given
redshift, but Equation~\ref{eq:GRB_rate} includes redshift evolution
of these properties for the relation GRB rate and the SFR. This
framework is equivalent to that of previous GRB studies such as
\citet{kistler09}. Recent studies of the nearby ($z<1$) GRB sample
suggest that these events may be biased towards lower-mass galaxies
(\citealt{levesque10a,svensson10}), possibly a result of either the standard
mass-metallicity relation for star-forming galaxies (e.g.,
\citealt{tremonti04}) or the fundamental metallicity relation
\citep{mannucci10}. The physical phenomenon driving this apparent bias
is not yet well understood (e.g.,
\citealt{levesque10a,levesque10b,kocevski11,niino11}). However if the
bias is related to a metallicity dependence, it should become less
important at high redshift \citep{fynbo08}. These unknown factors that
relate $\dot{\rho}_*$ to $\dot{n}_{GRB}$ complicate the use of GRBs as
tracers of SFR, however our analysis framework can be generalized to
include a luminosity/metallicity-dependent efficiency.

For LBG, we start by defining the integrated light above a given
luminosity $L$ as:
\begin{equation}\label{eq:l_int}
\mathcal{L}(L,z)=\int_{L}^{+\infty}\tilde{L}\phi(\tilde{L},z) d\tilde{L}. 
\end{equation}
From this, the relation with the SFR follows as:
\begin{equation}\label{eq:ll_i}
\mathcal{L}(0,z)=\eta_{LBG}(z)\dot{\rho_*}(z),
\end{equation}
where $\eta_{LBG}(z)$ is the conversion factor from observed luminosity
density to star formation rate (e.g., \citealt{madau96}). The star formation rate in galaxies with $L>L_{lim}$ is thus given as:
\begin{equation}\label{eq:rho_lim}
\dot{\rho}_{*}(L>L_{lim},z)=\frac{\mathcal{L}(L_{lim},z)}{\eta_{LBG}(z)}.
\end{equation}

SFR estimators from GRBs (probing all sites of star formation) and
LBGs (with observations at $L>L_{lim}$) can be derived
assuming models for $\varepsilon(z)$ and $\eta_{LBG}(z)$.  We
indicate these estimators as $\dot{\rho}_*^{(GRB)}(z)$ and
$\dot{\rho}^{(LBG)}_{*}(L>L_{lim},z)$.

The detected fraction of GRBs with host galaxies of luminosity $L>L_{lim}$ at
redshift $z$ is:
\begin{equation}\label{eq:fdet_intr}
f_{det}(L>L_{lim},z)=\frac{\dot{\rho}_{*}(L>L_{lim},z)}{\dot{\rho}_*(z)}\equiv\frac{\mathcal{L}(L_{lim},z)}{\mathcal{L}(0,z)}.
\end{equation}
Assuming that $\dot{\rho}_*^{(GRB)}(z)$ and $\dot{\rho}^{(LBG)}_{*}(L>L_{lim},z)$ are
unbiased estimators, we can rewrite Equation~\ref{eq:fdet_intr} as:
\begin{equation}\label{eq:fdet}
f_{det}(L>L_{lim},z)= \frac{\dot{\rho}^{(LBG)}_{*}(L>L_{lim},z)}{\dot{\rho}_*^{(GRB)}(z)}. 
\end{equation}

The fraction $f_{det}(L>L_{lim},z)$ allows us to measure the
relative amount of star formation in galaxies below the detection
threshold of the observations (Equation~\ref{eq:fdet_intr}). In fact,
under the assumption that there is no luminosity dependence on the
efficiency, GRBs are unbiased tracers of star formation at a given
redshift. Furthermore, the fraction of detected hosts gives the
relative amount of star formation present below $L_{lim}$, independent
of the specific value of $\varepsilon(z)$ and $\eta_{LBG}(z)$. From
$f_{det}(L>L_{lim},z)$ it is therefore immediate to derive the
integrated luminosity function at $L<L_{lim}$. 

To illustrate typical expected detection fractions of high-$z$ GRB
hosts as a function of redshift and survey depth, we show in
Figure~\ref{fig:fhost_vs_mlim} the predictions derived from our LF
model, assuming $M_{suppr}=-11$. Observations at the HUDF depth
($M_{lim}=-18$) are expected to detect $40-50\%$ of the GRB host halos
at $z\lesssim6$, but this fraction should decrease rapidly
at higher redshift. Although we resorted to a specific luminosity
function model to illustrate expected results, the relation in
Equation~\ref{eq:fdet_intr} is not model-dependent. Hence, the
observational determination of $f_{det}$ represents a powerful test to
determine the amount of star formation below the sensitivity of LBG
surveys.

With a model LF, it is also possible to go beyond the determination of
the integrated luminosity function at $L<L_{lim}$ and use $f_{det}$ to
constrain the luminosity scale at which galaxy formation is suppressed
and provide a test of galaxy formation theories and
simulations. Figure~\ref{fig:fhost_vs_msuppr} shows our model
predictions for a shallow ($M_{lim}=-20$) and deep ($M_{lim}=-18$)
survey as a function of $M_{suppr}$. Ideally, one would search for GRB
host galaxies at $z>6$, where $f_{det}$ is most sensitive to changes
in $M_{suppr}$. However, by using the larger sample of known GRBs at
$z\sim4-6$, it is also possible to constrain $M_{suppr}$ immediately
after the epoch of reionization, as discussed in
Section~\ref{sec:predictions}.

In addition, $f_{det}(L>L_{lim},z)$ provides a consistency check of
the model assumptions that lead to the determination of
$\dot{\rho}^{(LBG)}_{*}(L>L_{lim},z)$ and $\dot{\rho}_*^{(GRB)}(z)$.
The GRB efficiency $\varepsilon(z)$ is commonly 
described by Equation~\ref{eq:varepsilon}. Therefore,
Equation~\ref{eq:fdet} can be used to derive $\beta$. For example, if
$M_{suppr}\equiv M_{lim} =-17.7$ at $z_1=4$ [as suggested by the fact
that $\dot{\rho}^{(LBG)}_{*}(L>L_{lim},z_1)\approx
\dot{\rho}_*^{(GRB)}(z_1)$], and assuming that we have 
$M_{suppr}(z_1)=M_{suppr}(z_2)$ at $z_2\sim6>z_1$, then it follows that
$f_{det}(L>L_{lim},z_1)=f_{det}(L>L_{lim},z_2)$. From this, we derive:
\begin{equation}\label{eq:beta}
\beta =\log{\left(\frac{\dot{\rho}^{(LBG)}_{*}(L>L_{lim},z_2)}{\dot{\rho}^{(LBG)}_{*}(L>L_{lim},z_1)}\right)}/\log{\left(\frac{1+z_2}{1+z_1}  \right)}.
\end{equation}

\section{Predictions for high-$z$ searches of GRB host
  galaxies}\label{sec:predictions}

So far, studies of GRB-host galaxies have been primarily limited to
$z<4$ and are based on heterogeneous observations. Adopting a
compiled sample of 18 GRB hosts at $z<1$ from \citet{christensen04}
and \citet{levesque10a}, we find that nearby \emph{detected} hosts
have $\langle M_B\rangle=-19.3$. Similarly, \citet{svensson10} find that
GRB hosts at $z<1.2$ occur preferentially in small, relatively
low-mass 
galaxies. \citet{savaglio09} compile a larger sample of 45 GRBs
at $z<3.5$, with $\langle M_B\rangle =-20.3\pm0.5$, again for detected
hosts, as their catalog does not include GRBs for which only upper
photometric limits are available.

A systematic search of host galaxies at low redshift would help to
calibrate $\varepsilon(z)$ and construct the basis for comparison with
future detections of $z>4$ hosts at rest-frame UV and optical
wavelengths. The main problem currently is the limited sample size:
there are only 19 GRBs spectroscopically confirmed at $z>4$ and only
three at $z>6$. While this \emph{Letter} was under review, upper limits on
host galaxy luminosities for a sample of six $z>5$ GRBs observed with
\emph{HST} have been derived by \citet{tanvir12}. Still, a comprehensive
effort to detect these high-$z$ hosts is missing.

As proved by the \citet{tanvir12} sample, a systematic search is now
feasible for host galaxies in all known $z>4$ GRB to a magnitude limit
$M_{AB}\sim-18$. Using the {\it{Hubble}} Exposure Time Calculator, we
estimate that $6000~\mathrm{s}$ of integration time with WFC3 in
$F125W$ will reach a $2\sigma$ limit of $m_{AB}=28.5$ within a
diameter $d=0''.5$, corresponding to $M_{AB}\sim-18.0$, with a weak
dependence on the GRB redshift. Compared to LBG surveys, a GRB-host
survey has the key advantage that the position of the source is known
{\it a priori}. Therefore, detections at a lower signal-to-noise ratio
($S/N\sim2-3$) are still statistically significant. Based on the
surface density of sources in the {\it{Hubble}} Ultradeep Field Survey
\citep{bouwens11}, the chance of line-of-sight superposition with a
foreground faint source is low. However, data in a second band blueward
of the expected Lyman break would be useful to confirm that detected
hosts are at the GRB redshift. Ground-based observations using
10m-class telescopes with sensitivity comparable to \emph{HST} in the V and i
bands, provide a suitable alternative to search for $z\lesssim6$ hosts
and/or for deep observations in the second band (e.g., 
\citealt{basa12}).

\begin{figure}
\begin{center} 
\includegraphics[scale=0.4]{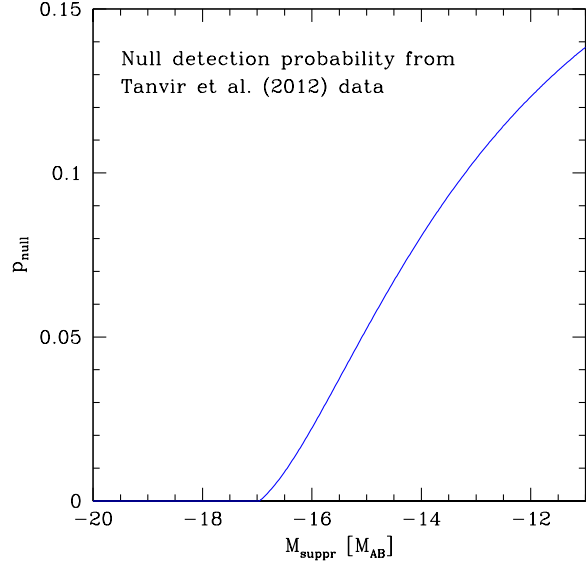}
\end{center}
\caption{Predicted probability of obtaining no detections of host
  galaxies in the \citet{tanvir12} sample ($6$ GRBs at $z>5$) based on
  our LF model, shown as function of suppression magnitude
  $M_{suppr}$. The non-detections constrain $M_{suppr}(z>5)>-15$ at
  $95\%$ confidence, demonstrating the importance of faint galaxies in
  the epoch of reionization.}\label{fig:tanvir_analysis}
\end{figure}

From our galaxy LF model, we expect $0.4\lesssim f_{det}\lesssim1$ at
$z\sim5$ depending on $M_{suppr}$. Therefore, a GRB host survey would
be expected to detect $\gtrsim8$ host galaxies at $4\leq z\leq6$. This
will measure $f_{det}$ to $\sim10\%$ accuracy, which will be
sufficient to determine whether $M_{suppr}$ is above or below
$M_{AB}=-15$ to high confidence level. To illustrate that this is a
feasible goal, we analyze in our framework the set of upper limits
$L_{lim, i}$ (with $i=1,6$) on host galaxy luminosity derived for the
$z>5$ GRBs sample in Table~2 of \citet{tanvir12}. With those limits we
construct the probability of null detection in their sample as a
function of the $M_{suppr}$ deriving $f_{det}$ from our LF model (see
Section~\ref{sec:model}):
\begin{equation}
p_{null}(M_{suppr}|\{L_{lim},z\}_i) = \Pi_i
\left[1-f_{det}(L>L_{lim, i},z_i|M_{suppr})\right]. 
\end{equation}
The results are shown in Figure~\ref{fig:tanvir_analysis} and
demonstrate that the absence of detected hosts constrains
$M_{suppr}>-15$ at 95\% confidence (and $M_{suppr}>-16.5$ at 99\%
confidence). This result is already a significant improvement upon the
limits inferred from current LBG surveys alone \citep[e.g.,][]{munoz11}. This technique also outperforms limits that can be
obtained in future LBG surveys with \emph{JWST}, which will reach
$M_{AB}\sim-16$ (however, \emph{JWST} will improve dramatically the
efficiency of the search for GRB hosts). In the future, additional
detections of GRBs at $z\sim8$ would best distinguish between models
with different suppression magnitudes, because $f_{det}(z=8)$ is very
sensitive to this quantity (see Figure~\ref{fig:fhost_vs_msuppr}).

\section{Conclusions}\label{sec:con}

\begin{figure}
\begin{center} 
\includegraphics[scale=0.35]{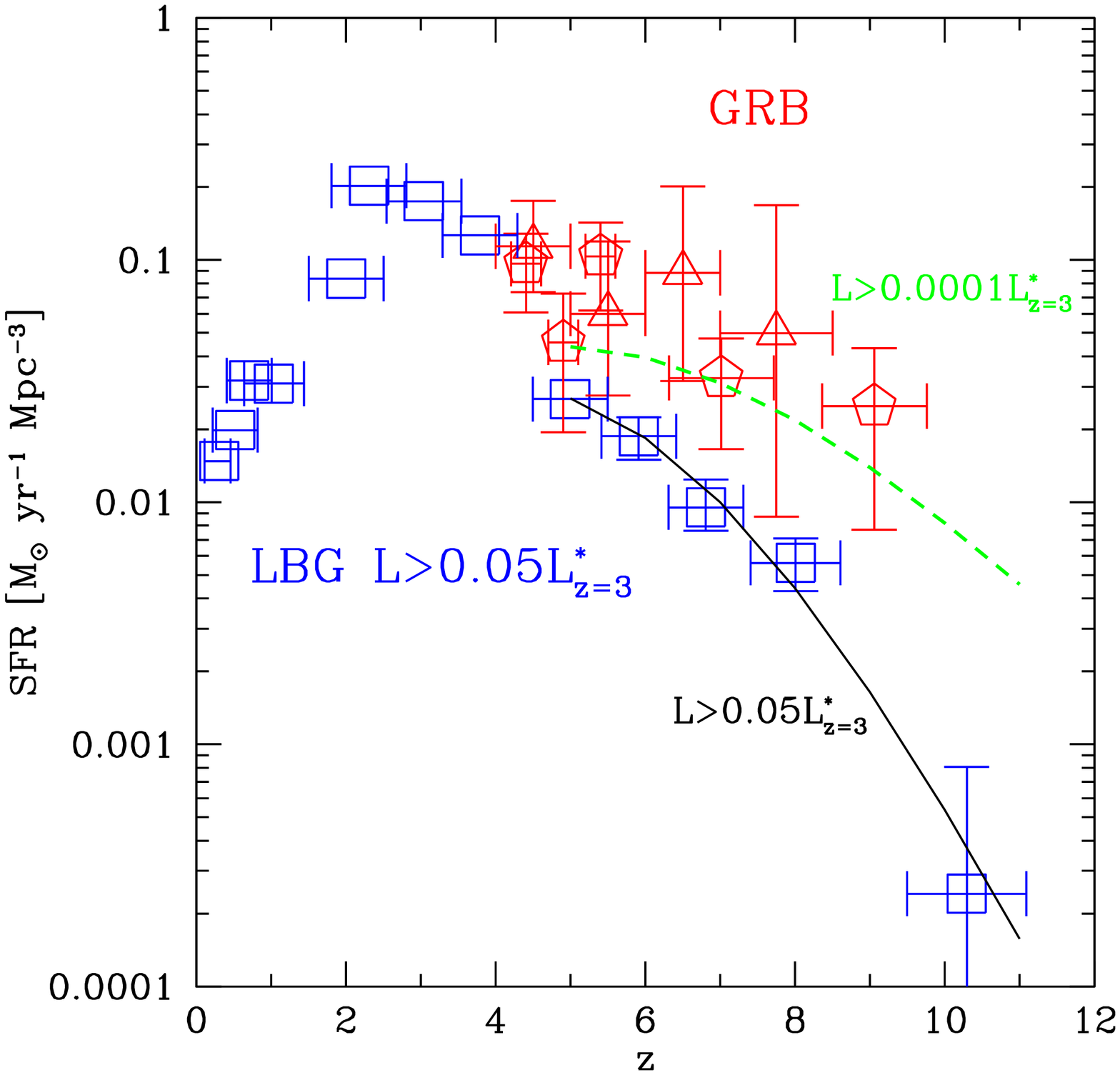}
\end{center}
\caption{Star formation rate as a function of redshift derived from
  surveys of Lyman break galaxies (LBG) by \citet{bouwens11} (blue squares)
  and gamma ray bursts (GRB) by \citet{kistler09} (red triangles) and
  \citet{robertson12} (``low-Z SFR'' model, red pentagons). The data
  from \citet{bouwens11} are integrated to a magnitude limit of
  $M_{AB}=-17.7$, corresponding to $L>0.05L^{*}_{z=3}$, and are
  corrected for dust extinction. GRB surveys probe all sites of star
  formation, including galaxies below the current detection limit for
  LBG surveys. The black-solid and green-dashed lines show the SFR
  predicted by the high-$z$ LF model of \citet{trenti10a} integrated to
  the same limit as LBG surveys (black) and down to the limit of
  H~I cooling halos ($M_{AB}=-11$, green).}\label{fig:sfr}
\end{figure}

In this {\it{Letter}} we discuss the relation between SFR estimates
from GRB and LBG surveys. The GRB-inferred rate is higher than that of
LBGs at $z>4$ (see Figure~\ref{fig:sfr}), suggesting that significant
star formation takes place in galaxies below the LBG detection
limit. The difference between the two approaches can be qualitatively
explained by the model of galaxy formation based on the dark-matter
halo evolution developed by \citet{trenti10a}, under the assumption
that star formation proceeds down to the limit of H~I cooling
($T_{vir}\gtrsim2\times10^4~\mathrm{K}$ or
$M_{DM}\gtrsim2\times10^8~\mathrm{M_{\sun}}$). In this model, the
assembly of galaxies is linked to the growth of their dark-matter
halos. By construction, our model is consistent with the build-up of
stellar mass density, unaffected by the over-production of stellar
mass at $z=4$ inferred from the GRB-SFR by \citet{robertson12}. In
this respect, it is interesting that our SFR predictions in
Figure~\ref{fig:sfr} are systematically $\sim1\sigma$ lower than the
datapoints from GRB observations. Our findings suggest a mild
systematic over-estimation of the SFR derived from GRBs at $z>4$, as
concluded by \citet{robertson12} and by \citet{choi12}.

To gain further insight on the relation between the SFR estimates from
GRB and LBG surveys, we introduced the idea that the fraction
$f_{det}(L>L_{lim},z)$ of detected GRB hosts in a survey with
$L>L_{lim}$ provides an unbiased estimator of star formation. The
relative amount of star formation in undetected faint galaxies
(Equation~\ref{eq:fdet_intr}) can quantity the role played by galaxies
during hydrogen ionization. Starting from $f_{det}(L>L_{lim},z)$ and
using a LF model, it is possible to determine the scale at which
galaxy formation is suppressed at low masses. Furthermore,
$f_{det}(L>L_{lim},z)$ can be used to measure variations with redshift
in the evolution of the GRB efficiency per unit stellar mass
(Equations~\ref{eq:GRB_rate} and \ref{eq:beta}), a key quantity to
understand the production of GRBs across cosmic time.

\begin{figure}
\begin{center} 
\includegraphics[scale=0.35]{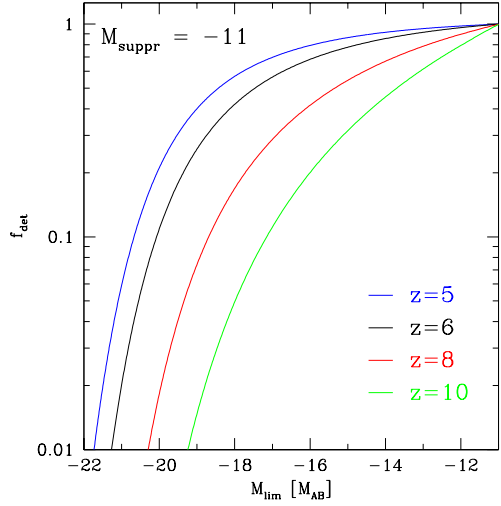}
\end{center}
\caption{Predicted fraction of GRB-host-galaxy detections
  $f_{det}(L>L_{lim},z)$, as a function of the limiting magnitude of
  the survey [$M_{lim}=-2.5\log_{10}{(L_{lim})}$] for a model with
  star formation in galaxies with absolute AB magnitude
  $M<M_{suppr}=-11$. We predict a significant detection fraction for
  deep surveys with $M_{lim}\sim-18$ [$f_{det}(L>L_{lim},z)\gtrsim0.4$] at
  $z\lesssim6$, and a decrease at higher
  redshift.}\label{fig:fhost_vs_mlim}
\end{figure}

\begin{figure} 
\includegraphics[scale=0.41]{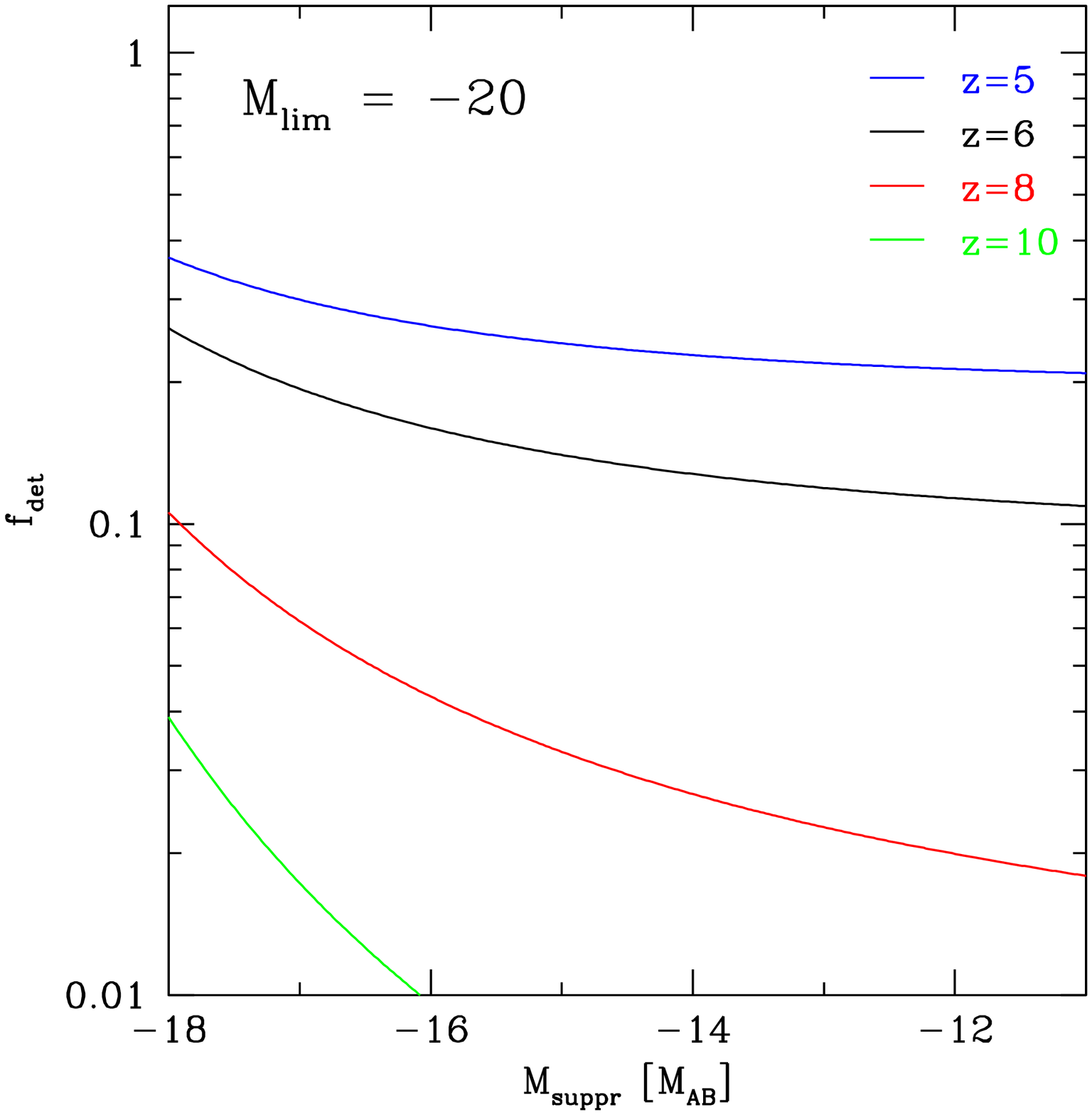}
\includegraphics[scale=0.41]{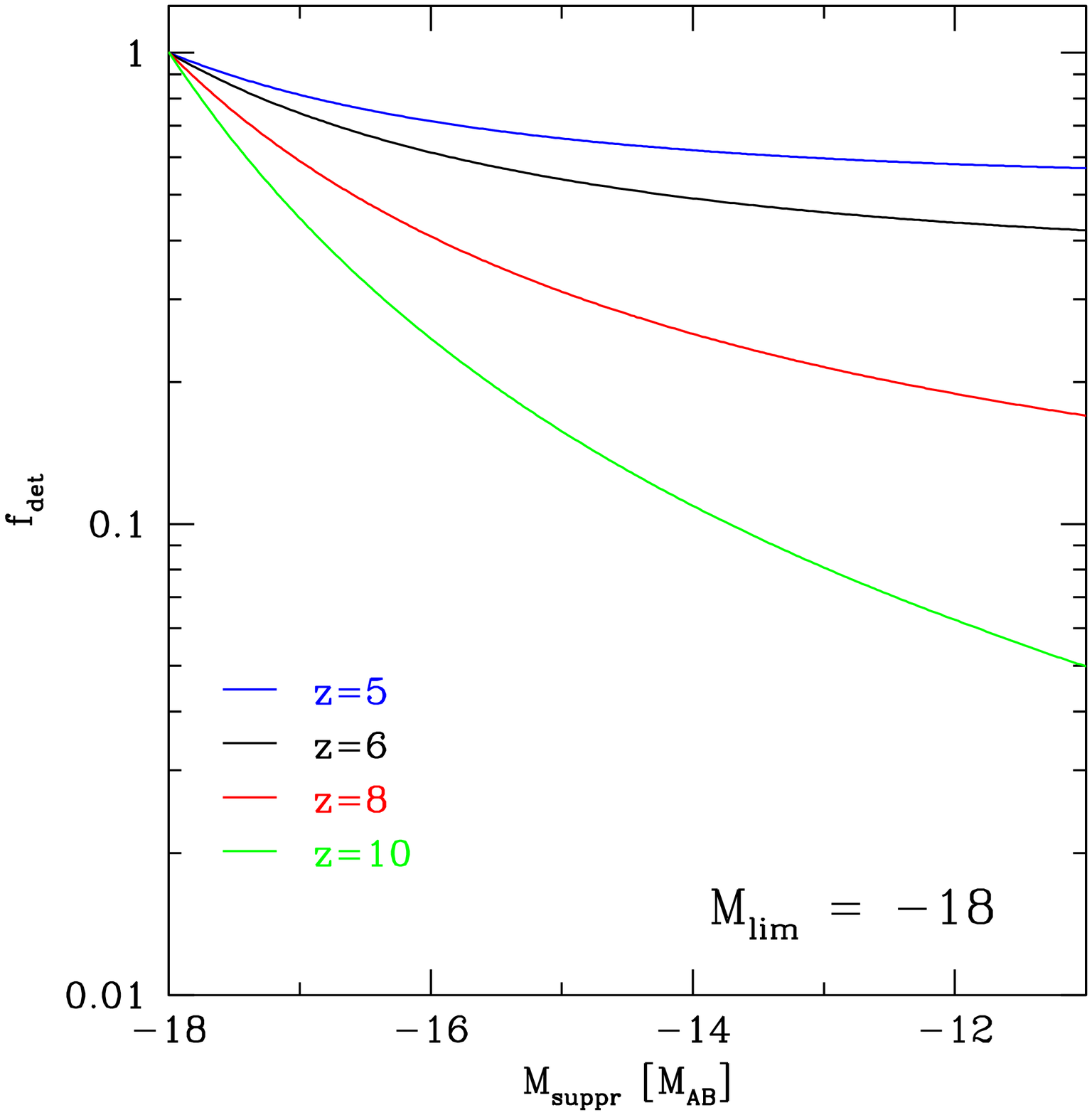}
\caption{Predicted fraction of GRB host galaxy detections
  $f_{det}(L>L_{lim},z)$, as a function of suppression magnitude
  $M_{suppr}$ for a shallow survey ($M_{lim}=-20$, left panel) and for
  a deep survey ($M_{lim}=-18$, right panel). Measuring $f_{det}$
  allows one to determine $M_{suppr}$.}\label{fig:fhost_vs_msuppr}
\end{figure}

Based on our specific galaxy formation model, we made predictions of
the expected detection fraction of GRB hosts at high-$z$
(Figures~\ref{fig:fhost_vs_mlim}-\ref{fig:fhost_vs_msuppr}). We expect
that $\sim50\%$ of the GRB hosts could be detected at $z\sim5$,
followed by a sharp drop at higher redshift because of a steep
faint-end slope of the galaxy LF. In general, the more that star
formation is dominated by low-mass, low-luminosity halos, the smaller
the detected host fraction. Our modeling is consistent with the
non-detection of GRB hosts for the six highest-redshift GRBs known to
date \citep{tanvir12}. The analysis of their limits in our framework
allows us to constrain $M_{suppr}>-15$ at 95\% confidence (and
$M_{suppr}>-16.5$ at 99\% confidence), demonstrating that the majority
of ionizing photons at $z\gtrsim6$ were produced in small,
low-luminosity galaxies.  A systematic search for GRB hosts down to
faint luminosity limits ($M_{AB}\sim-18$) for all known $z\geq4$ GRBs
would improve the limits on $M_{suppr}$ and could provide the
definitive proof that the faintest galaxies are the agents of hydrogen
reionization.

\acknowledgements 

We thank Richard Ellis, Ken Nagamine, Brant Robertson,
Nial Tanvir, Bing Zhang, and the referee for helpful comments and
suggestions. This work is supported in part by grants NASA ATP-NNX07AG77G,
NSF AST-0707474, NSF AST-1009396, NSF PHY11-25915, STScI-HST-GO-11700. 






\end{document}